\documentclass[aps,prd,preprint,nofootinbib]{revtex4}
\usepackage{epsfig}
\usepackage{graphicx}
\newcommand{\bea}{\begin{eqnarray}}
\newcommand{\eea}{\end{eqnarray}}
\newcommand{\be}{\begin{equation}}
\newcommand{\ee}{\end{equation}}
\newcommand{\ba}{\begin{array}{l}}
\newcommand{\ea}{\end{array}}

\newcommand{\wt}{\widetilde}
\newcommand{\wth}{\widetilde{H}}

\newcommand{\sto}{\widetilde{t}_1}

\newcommand{\ra}{\rightarrow}
\newcommand{\e}{\epsilon}

\begin{document}
%%====================   Pretext     ========================
\title{Fragmentation Function and Hadronic Production of the
Heavy Supersymmetric Hadrons}

\author{Chao-Hsi Chang$^{1,2,4}$\footnote{email:
zhangzx@itp.ac.cn}, Jiao-Kai Chen$^{2,3}$\footnote{email:
jkch@mail.haust.edu.cn}, Zhen-Yun Fang$^{4}$, Bing-Quan Hu$^{4}$,
Xing-Gang Wu$^{2,4}$\footnote{email: wuxg@itp.ac.cn}}

\address{$^1$CCAST (World Laboratory), P.O.Box 8730, Beijing 100080,
China.\\
$^2$Institute of Theoretical Physics, Chinese Academy of Sciences,
P.O.Box 2735, Beijing 100080, China.\\
$^3$ Department of Mathematics and Physics, Henan University of
Science and Technology, Luoyang, China, 471003.\\
$^4$Department of Physics, Chongqing University, Chongqing, 400044,
China.}

\begin{abstract}
The light top-squark $\sto$ may be the lightest squark and its
lifetime may be `long enough' in a kind of SUSY models which have
not been ruled out yet experimentally, so colorless `supersymmetric
hadrons (superhadrons)' $(\sto \bar{q})$ ($q$ is a quark except
$t$-quark) may be formed as long as the light top-squark $\sto$ can
be produced. Fragmentation function of $\sto$ to heavy
`supersymmetric hadrons (superhadrons)' $(\sto \bar{Q})$
($\bar{Q}=\bar{c}$ or $\bar{b}$) and the hadronic production of the
superhadrons are investigated quantitatively. The fragmentation
function is calculated precisely. Due to the difference in spin of
the SUSY component, the asymptotic behavior of the fragmentation
function is different from those of the existent ones. The
fragmentation function is also applied to compute the production of
heavy superhadrons at hadronic colliders Tevatron and LHC under the
so-called fragmentation approach. The resultant cross-section for
the heavy superhadrons is too small to observe at Tevatron, but
great enough at LHC, even when all the relevant parameters in the
SUSY models are taken within the favored region for the heavy
superhadrons. The production of `light superhadrons' $(\sto
\bar{q})$ ($q=u, d, s$) is also roughly estimated. It is pointed out
that the production cross-sections of the light superhadrons $(\sto
\bar{q})$ may be much greater than those of the heavy superhadrons,
so that even at Tevatron the light superhadrons may be produced in
great quantities.\\



\end{abstract}

\maketitle

%%%====================  Text     =======================
\section{Introduction}

Supersymmetry (SUSY) is one of the most appealing extensions of the
Standard Model (SM)\cite{bagger,bagger1,nilles,nilles1,haber}.
Whereas without knowing the realistic SUSY breaking mechanism, even
in the minimum supersymmetry extension of the Standard Model (MSSM),
there are too many parameters in the SUSY sector, which need to be
fixed via experimental measurements. If the MSSM is rooted in the
`minimum supergravity model' (mSUGRA), the numbers of the
independent parameters can be deducted a lot, but there are still
many un-fixed parameters\cite{haber,susy1,Ellis1}. Therefore, the
spectrum of the SUSY sector in SUSY models still is a quite open
problem\footnote{In fact, all of the available indications on the
masses of the SUSY partners are abstracted from experimental
measurements and/or astro-observations under assumptions (not direct
measurements), so one should consider them only as references.}.

For some kinds of SUSY models and by choosing un-fixed parameters
from the region which has not been ruled out yet, it is not
difficult to realize a general feature of the two mass eigenstates
$\sto$ and $\widetilde{t}_2$ for the SUSY partners of top-quark
(top-squark $\widetilde{t}_L$ and $\widetilde{t}_R$) such as that
the comparatively lighter one $\sto$ is the lightest squark, and the
$\sto$'s lifetime is so `long', that its width is less than
$\Lambda_{QCD}$\cite{s-ellis,s-ellis1,hikasa,dj,dj1,dj2,s-benaker,datta}.
In this case, $\sto$ may form various colorless hadrons i.e. the
superhadrons by QCD interaction, which consist of $\sto$ and
$\bar{q}$ (here $q=u,d,c,s,b$). On the other hand, the direct
experimental searching for the SUSY partners may only set a lower
bound of $\sto$ mass as $m_{\sto}\geq 100$
GeV\cite{LEP,Teva,Teva1}\footnote{For a summary see \cite{LEP}.}
(even lower as $75$ GeV\cite{datta}). Therefore in the paper, we
would like to focus our attention on the consequences for the
possible features of $\sto$. Namely, we shall assume that $\sto$,
the SUSY partners of top-quark, is not very heavy e.g.
$m_{\sto}\simeq 120\sim 150$ GeV and has a `quite long' lifetime
$\Gamma_{\sto} \leq \Lambda_{QCD}$, that $\sto$ (after producing and
before decaying) has chances to form colorless superhadrons
($\sto\bar{q}$)\footnote{This kind of superhadrons are bound states
of a quark (anti-quark) and an anti-squark (squark), or two gluinos,
or two quarks (antiquarks) and a squark (anti-squark), etc. All of
them are colorless and bound via strong interaction
(QCD)\cite{nappi,chcui,chcui1}.}. Moreover we think that the squark
(anti-squark) in the superhadrons is a scalar, which is different
from a quark in `common' hadrons, hence with such a scalar component
the study of the superhadrons is also very interesting from the
bound state point of view.

There was a remarkable progress in nineties of the last century in
perturbative QCD (pQCD) for double heavy meson studies, i.e., it was
realized that the fragmentation function and the production of a
double heavy meson such as $B_c$ and $\eta_c, J/\psi$ can be
reliably computed in terms of pQCD and the wavefunction derived from
the potential
model\cite{chang92,chang921,chang922,chang923,braatenc,braatenc1,prod3}\footnote{Exactly
to say, here we mean color-singlet mechanism only i.e. only the
color-singlet component of the concerned double heavy meson, which
is the biggest in Fock space expansion, is taken into account in
calculating the fragmentation function and production as well. While
since the color-octet matrix element appearing in the formulation
for color-octet mechanism production (fragmentation function) cannot
be calculated theoretically so far, so the color-octet mechanism is
not the case.}, and further progress in formulating the problem
under the framework of the effective theory: non-relativistic QCD
(NRQCD) \cite{nrqcd,nrqcd1} was made a couple years late. We should
note here that before
Refs.\cite{chang92,chang921,chang922,chang923,braatenc,braatenc1,prod3}
there were papers \cite{be,ji}. The inclusive production of a meson
\cite{be,ji} and the fragmentation functions of a parton into a
meson \cite{ji} were calculated. But the authors of \cite{ji}
precisely claimed that their calculations might be extended to the
cases for the heavy-light mesons, such as $D, B$ etc.. In fact, the
claim is incorrect and misleading in the key point on the
theoretical calculability of the production and the fragmentation
functions\footnote{To present the calculations of the fragmentation
functions, we would like also to remind here that there are some
substantial contributions from the phase-space integration which
might be missed if enough care were not paid. In fact, as pointed
out in \cite{chang921}, they were missed in \cite{ji} indeed.}.
Since heavy-light mesons contain a light quark, and the light quark
creation involving in the fragmentation function is
non-perturbative, so it cannot further be factorized out a hard
factor as that in the case of the double heavy mesons. It is just
the reason why, similar to the case for double heavy mesons, we
expect that only the `heavy superhadrons' ($\widetilde{t}_1
\bar{Q}$) (but not the `light superhadrons' ($\widetilde{t}_1
\bar{q}$)), where $Q$ ($q$) denotes a heavy quark, $c$ or $b$ (a
light q1uark, $u$ or $d$ or $s$), and their inclusive production and
fragmentation functions may be calculated reliably. The
fragmentation functions to the `heavy superhadrons' ($\sto \bar{Q}$)
can be simply attributed to a wave function of potential model and a
pQCD calculable factor as in the case of the double heavy mesons.

For the `light superhadrons' ($\sto \bar{q}$) where $q$ indicates a
light quark: $u$ or $d$ or $s$, due to the non-perturbative nature
for producing the involved light quark $q$, the `story' about the
calculation of the fragmentation function is very different, i.e.,
the fragmentation function cannot be attributed to a wave function
and a hard factor of pQCD. Due to the non-perturbative QCD effects
in the fragmentation function of the `light-heavy mesons' such as
$B$, $D$ and etc, so far practically the way to obtained the
fragmentation functions of the `light-heavy meson' is that they are
formulated in terms of theoretical considerations and
parametrization first, and then the parameters in the formulation
are fixed via experimental measurement(s). With the fragmentation
functions the production cross-section of a `light-heavy meson', in
experiences, generally is greater than that of the respective double
heavy meson (the $q$ quark in `light-heavy meson' is replaced by a
charm quark $c$) by a factor $10^{3\sim 4}$. Since there is no
experimental observation on superhadrons, so we cannot play the same
way for the `light superhadrons' as `light-heavy mesons' at all.
Alternatively, as a magnitude order estimate, we expect that the
fragmentation function and the production of the light superhadrons
($\sto \bar{q}$) are also greater than that of the respective heavy
superhadrons ($\sto \bar{Q}$) by a factor $10^{3\sim 4}$, no matter
how heavy $\sto$ is, that is very similar to the case of a double
heavy meson vs a heavy-light meson. Thus, based on the quantitative
computation of the fragmentation function and the production of the
heavy superhadrons, we simply extend the results of the the
production to the light superhadrons at the end of the paper by
referring the cases of the `double heavy mesons' vs the `light-heavy
mesons' experientially. For convenience, later on we will denote
($\sto \bar{Q}$) as $\wth$ throughout the paper.

Since the spectrum and the wave function accordingly of a double
heavy-quark binding system i.e. a system of a heavy quark and a
heavy anti-quark system, ($Q'\bar{Q}$), can be obtained
theoretically in terms of non-relativistic potential model inspired
on QCD quite well, so the `heavy superhadrons' $\wth$, as the double
heavy system ($Q'\bar{Q}$), may also be depicted by the
non-relativistic potential model as long as the difference in spin
is taken into account carefully\cite{chcui,chcui1}. Therefore, there
is no problem to obtain the wave functions of `heavy superhadrons'
$\wth$ that appear in the fragmentation functions.

In the literature, there are two approaches for estimating the
direct production of a double heavy meson in NRQCD framework: the
`fragmentation approach' vs the complete `lowest-order-calculation'
approach. It is known that, of the two approaches, the former is
much simpler than the latter one in computation, but the former is
`good' only in the region where the transverse momentum of the
produced double heavy meson is large ($p_T\gtrsim
15$GeV)\cite{prod4,prod41,prod42}. The situation for the production
of the heavy superhadrons is similar to the cases of the double
heavy mesons, so of the two approaches, we adopt the fragmentation
approach when estimating the production of the superhadrons for
rough estimation.

This paper is organized as follows: In Section II, we show how to
derive the fragmentation function of the lightest top-squark $\sto$
to the `heavy superhadrons' $\wth$, and try to present the obtained
results i.e. its general features properly. In Section III, we
compute the cross sections for hadronic production of the
superhadrons at colliders Tevatron and LHC in terms of the so-called
fragmentation approach. The Section IV is devoted to the discussions
and conclusions.

%%%%====================================================================
\section{Fragmentation function of the light top-squark $\sto$ to
the heavy superhadrons $\wth$}

In this paper we adopt the fragmentation approach to estimate the
$\wth$ production, and in the present section, we are computing the
fragmentation function first, that is one of the key factors of the
fragmentation approach.

According to pQCD, with leading logarithmic (LL) terms being summed
up, a fragmentation function of a `parton' $i$ to a heavy
superhadron $\wth$ is depicted by DGLAP Equation as
below\cite{AP,AP1,AP2}:
\be { \frac{dD_{i \ra \wt{H}}(z,Q^2)}{d\tau} =
\sum_j\frac{\alpha_s(Q^2)}{2\pi}\int_z^1 {dy \over y} \; P_{i \ra
j}(z/y)
    \; D_{j \ra \wt{H}}(y,Q^2) \;,
} \label{evol} \ee
where $\tau=\log(Q^2/\Lambda_{QCD}^2)$, $P_{i\ra j}(x)$ is the
splitting function. For example, the splitting function for the
supersymmetric top-squark \cite{splitting,splitting1} reads
\begin{displaymath}
 P_{\sto \ra \sto g}\left(x\right)
 =\frac{4}{3}\left[\frac{1+x^2}{(1-x)_+}-(1-x)+\delta(1-x)\right].
\end{displaymath}

Since Eq.(\ref{evol}) is an integro-differential equation, so to
have definite solution, a `boundary (initial) condition' for the
equation i.e. $D_{j \ra \wt{H}}(z,Q_0)$, the fragmentation function
at the energy scale $Q_0\simeq m_{\tilde{t}_1}$, is needed. Now the
task is to obtain the boundary condition. Fortunately, the boundary
condition for heavy superhadron $\wth$ can be derived in terms of
pQCD and the relevant wave function precisely as double heavy
mesons\cite{chang92,chang921,chang922,chang923,braatenc,braatenc1,prod3}.
Hereafter, to simplify our notation, we shall always use $D_{j \ra
\wt{H}}(z)$ instead of $D_{j \ra \wt{H}}(z,Q_0)$.

Since the fragmentation functions are universal by definition, i.e.
they are independent of the concrete process, so for the `boundary
(initial) condition' of the fragmentation function of the light
top-squark $\sto$, we would like to choose a relevant simple process
to calculate the `boundary condition' $D_{\sto \ra \wt{H}}(z)$. In
order to simplify the derivation as much as possible, we furthermore
assume a fictitious $``Z"$, which, except the mass, has the same
properties as that of the physical $Z$ boson. The fictitious $``Z"$
has such a great mass that it may decay to a pair of $\sto$ and
$\bar{\sto}$.

According to pQCD factorization theorem, the differential width for
the fictitious $``Z"$ decaying to $\wth$ may be factorized as
\be d\Gamma (``Z" \ra \wth + X) = \int_{0}^{1} dz\;
d\widehat{\Gamma}(``Z"\ra \sto  + \bar{\sto},\mu_{f})
     D_{\sto \ra \wth}(z,\mu_{f}) ,
 \label{facZ}
\ee
where $z=\frac{2E}{\sqrt{S_{eff}}}$ and $\mu_f$ is the energy scale
for factorization. By definition, $D_{\sto \ra \wth}(z,\mu_{f})$ is
the fragmentation function, which represents the probability of
$\sto$ fragmenting into the superhadron with energy fraction $z$.

%%%-----------------------------
%%frag function comparision
\begin{figure}
\centering
\includegraphics[scale=0.6]{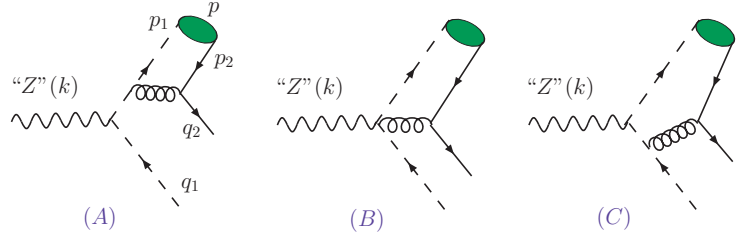}
\caption{The Feynman diagrams for the fictitious particle $``Z" (k)$
decaying into a superhadron $\wth (p)$, $\bar{\sto} (q_1)$ and a
bottom quark $b(q_2)$ (or a charm quark $c(q_2)$). (A), (B), (C) are
the Feynman diagrams with the hard virtual-gluon attached to the
light top-squark $\sto$ and its anti-particle $\bar{\sto}$ in
different ways.}\label{zstopf}
\end{figure}

To calculate $D_{\sto \ra \wth}(z)$, let us calculate the process
$``Z" \ra \wth + X$ precisely. Of the lowest-order pQCD calculation,
the Feynman diagrams for the inclusive decay of the fictitious
particle $``Z"$ into a $({1\over 2})^-$ superhadron $\wth$ (with
mass $M$), $``Z" \ra \wth + X$, are described by the three diagrams
(A), (B) and (C) as shown in FIG.\ref{zstopf}. The intermediate
gluon in each of the Feynman diagrams should ensure the production
of a heavy quark-antiquark pair, so its momentum squared should be
bigger than $(p_2+q_2)^2\geq 4m_Q^2\gg \Lambda_{QCD}^2$, thus pQCD
calculation and the factorization theorem are reliable. The
corresponding amplitudes are
\begin{eqnarray}\label{eq00}
M&=&\frac{4g^{2}_{s}}{3\sqrt{3}}\frac{gc_{11}} {cos\theta_{W}} \int
d^{4}q \;\mathrm{Tr}
\left\{\chi^{(1/2)^{-}}(p,q)(h^{\mu}_{A}+h^{\mu}_{B} +h^{\mu}_{C})
\frac{G_{\mu\nu}}{(p_{2}+q_{2})^{2}}\overline{u}(q_{2})\gamma^{\nu}\right\}\nonumber\\
&\simeq&\frac{4g^{2}_{s}}{3\sqrt{3}}\frac{gc_{11}}
{cos\theta_{W}}g_{B}(h^{\mu}_{A}+h^{\mu}_{B}+h^{\mu}_{C})
\frac{G_{\mu\nu}}{(p_{2}+q_{2})^{2}}L^{\nu},
\end{eqnarray}
with
\begin{displaymath}
h^{\mu}_{A} =\frac{(p_{1}+k-q_{1})^{\mu}}{(k-q_{1})^{2}
-m_{\sto}^{2}}(k-2q_{1})\cdot\epsilon,\;\;\; h^{\mu}_{B} =
-2\epsilon^{\mu},\;\;\; h^{\mu}_{C} = \frac{(p_{1}-k-q_{1})^{\mu}}
{(k-p_{1})^{2}-m_{\sto}^{2}} (2p_{1}-k)\cdot\epsilon
\end{displaymath}
and
\begin{displaymath}
\displaystyle L^{\nu}=\overline{u}(q_{2})\gamma^{\nu}v(p),\;\;\;
G_{\mu\nu}= g_{\mu\nu}-\frac{q_{1\mu}(q_{2}+p_{2})_{\nu}
+q_{1\nu}(q_{2}+ p_{2})_{\mu}}{q_{1}\cdot(q_{2}+p_{2})},\;\;\;
g_B=\frac{\phi(0)}{ 2\sqrt{m_{\sto}}}\,,
\end{displaymath}
where $k$, $p_{1}$, $p_{2}$, $q_{1}$ and $q_{2}$ are the four
momenta of $``Z"$, top-squark $\sto$, antiquark $\bar{Q}$ ($\bar{b}$
or $\bar{c}$), top anti-squark $\bar{\sto}$ and quark $Q$ ($b$ or
$c$) respectively. $p$ is the four momentum of $\wth$ and $q$ is the
relative momentum between the two constituents inside $\wth$, so we
have $p=p_{1}+p_{2}$, $q=\alpha_{2}p_1-\alpha_{1}p_2$ with
$\alpha_1=\frac{m_1}{m_1+m_2}, \alpha_2=\frac{m_2}{m_1+m_2}$, where
$m_1$ and $m_2$ are masses of $\sto$, $\bar{Q}$ respectively.
$\phi(0)$ is the wave function at origin for the superhadron $\wth$.
If neglecting the $q$-dependence of the integrand of
Eq.(\ref{eq00}), which can be considered as the lowest term in the
expansion of $q$ for the integrand, all the non-perturbative effects
can be attributed into the wave function at origin after doing the
integration over $q$. $\epsilon$ is the polarization vector for the
fictitious particle $``Z"$.
$c_{11}=I_{3L}cos^{2}\theta_{\widetilde{q}} -e_{q}sin^{2}\theta_{W}$
\cite{kraml} \footnote{$c_{11}$ is a factor in the effective
coupling $``Z"-\sto-\bar{\sto}$, which will not appear in the final
result of the fragmentation function because of cancelation between
the numerator and denominator in the computation formula
Eq.(\ref{sec:frag}).}. For convenience, the variables:
$S_{eff}=k^{2}$, $x=\frac{2p\cdot k}{S_{eff}}$, $y=\frac{2q_{1}\cdot
k}{S_{eff}}$, $z=\frac{2q_{2}\cdot k}{S_{eff}}$, $M=m_{\sto}+m_{Q}$
and $d=\frac{M}{\sqrt{S_{eff}}}$ are introduced. Keeping the leading
term for $d^{2}$, the maximum and minimum values of $y$ are
$y_{max}=1-\frac{d^{2}(1-\alpha_{1}x)^{2}}{x(1-x)}$,
$y_{min}=1-x+\frac{d^{2}(1-x+\alpha_{1}x)^{2}}{x(1-x)}$. In the
calculation the axial gauge $n_\mu=q_{1\mu}$ is adopted. Under the
axial gauge, it can be found that only the two amplitudes $M_{A}$
and $M_{B}$, which correspond to the first two Feynman diagrams in
FIG.\ref{zstopf}, have contributions to the fragmentation function.

According to the factorization Eq.(\ref{facZ}), the fragmentation
function versus $z$ at energy scale $Q_0$ can be derived by dividing
the differential decay width with $\Gamma_0$:
\bea\label{sec:frag}
 D_{\wt{t}_1\ra\wth}(z)&=&\frac{1}{\Gamma_0}
 \frac{d\Gamma}{dz},
 \eea
where $\Gamma_0$ is the decay width for the fictitious particle
$``Z"$ decaying into top-squark $\sto$ and top anti-squark
$\bar{\sto}$. Thus the result for the fragmentation function may be
presented as follows:
\begin{eqnarray}\label{Dst}
D_{\sto\ra\wth}(z)=F_{\sto}\cdot f_{\sto}(z),
\end{eqnarray}
where
\begin{eqnarray}\label{stt}
&\displaystyle
F_{\sto}=\frac{16\alpha_{s}(4m^2_{Q})|\phi(0)|^{2}}{27\pi
m_Q^2m_{\sto}},\nonumber\\
&\displaystyle f_{\sto}(z)=\frac{1}{6}
\frac{(1-z)^{2}z^{2}}{(1-\alpha_{1}z)^{6}}\cdot\left[2\alpha^2_{1}(z-4)z
+\alpha^3_{1}(3\alpha_{1}z-2z+2)z
+3\alpha^2_{1}-6\alpha_{1}+6\right].
\end{eqnarray}

\begin{table}
\begin{center}
\caption{The wavefunction at origin $\phi(0)$ for (${\sto\bar{b}}$)
or (${\sto\bar{c}}$). Here the needed parameters are set: $m_b=5.18$
GeV and $m_c=1.84$ GeV. }\vspace{2mm}
\begin{tabular}{|c|c|c|}
\hline ~~~$m_{\sto}$~~~ & ~~~$120$~GeV~~~ & ~~~$150$~GeV ~~~\\
\hline\hline $\phi(0)_{(\sto\bar{b})}\left[(GeV)^{3/2}\right]$& $2.502$ & $2.530$ \\
\hline $\phi(0)_{(\sto\bar{c})}\left[(GeV)^{3/2}\right]$& $0.693$ & $0.695$\\
\hline\hline
\end{tabular}
\label{wavezero}
\end{center}
\end{table}

At preset, there is no experiment data for the superhadron at all,
so we adopt potential model with the Cornell potential
$\left(-\frac{\kappa}{r}+ {r\over {a^2}}\right)$ to estimate the
wave function at origin, $\phi(0)$. For definiteness, we assume that
the potential of the heavy scalar-antiquark binding system is the
same as that of double heavy quark-antiquark systems. The relevant
parameters are taken as: $\kappa=0.52$, $a=2.34$ GeV$^{-1}$
\cite{eichten}, $m_{\sto}=120$ or $150$ GeV, $m_b=5.18$ GeV and $
m_c=1.84$ GeV. The corresponding wave functions at origin $\phi(0)$
for the systems ($\sto\bar{b}$) and ($\sto\bar{c}$) are listed in
TAB.\ref{wavezero}.

The fragmentation function $D_{\sto\ra\wth}(z)$ obtained, see
Eq.(\ref{Dst}), is just a boundary condition for the DALAP evolution
equation Eq.(\ref{evol}). Solving the DGLAP equation, one may obtain
the fragmentation function with the energy-scale evolution to $Q^2$.
The relevant Feynman diagrams for the boundary condition $D_{j \ra
\wt{H}}(z)$ with $(j=q, \bar{q}, g)$ are of higher order in
$\alpha_s$ than the Feynman diagrams for $D_{\sto \ra \wt{H}}(z)$,
see FIG.\ref{zstopf}\footnote{For $D_{g \ra \wt{H}}(z)$, the
relevant part of the Feynman diagrams must have one more strong
coupling vertex ($g\ra \sto\bar{\sto}$) in $\alpha_s$ than the
relevant part $\sto\ra \wth$ in FIG.\ref{zstopf}, and for
$D_{q(\bar{q}) \ra \wt{H}(z)}$ the relevant Feynman diagrams must
have two more strong coupling vertex $q\ra qg$ and $g\ra
\sto\bar{\sto}$ in $\alpha_s$ than those $\sto\ra \wth$.}, therefore
only the case with $i, j=\sto$ shall be taken into account in
solving Eq.(\ref{evol}), so as to meet the LL approximation
criterion.

We solve Eq.(\ref{evol}) with the method developed by Field
\cite{field}. When $Q^2\gg m_{\sto}^2$, according to the Field's
method, we have the following solution:
\begin{eqnarray}\label{ff-appr}
&\displaystyle
D_{\sto\ra\wt{H}}(z,Q^2)=\wt{D}_{\sto\ra\wt{H}}(\frac{8}{3},z,Q^2) +
\kappa\int_z^1{dy\over
y}\wt{D}_{\sto\ra\wt{H}}(\frac{8}{3},z/y,Q^2)P_{\Delta}(y)
\nonumber\\
& \displaystyle + \kappa\int_z^1{dy\over
y}\wt{\wt{D}}_{g\ra\wt{H}}(z/y,Q^2)P_{\sto\ra g}(y)
+O(\kappa^2)\,,\nonumber\\
&\displaystyle D_{g\ra\wt{H}}(z,Q^2)=\wt{D}_{g\ra\wt{H}}(6,z,Q^2) +
\kappa\int_z^1{dy\over y}\wt{D}_{g\ra\wt{H}}(6,z/y,Q^2)P_{\Delta
g}(y)
\nonumber\\
& \displaystyle + \kappa\int_z^1{dy\over
y}\wt{\wt{D}}_{\sto\ra\wt{H}}(z/y,Q^2)P_{g\ra \sto}(y)
+O(\kappa^2)\,,
\end{eqnarray}
with
\begin{eqnarray}
& \displaystyle P_{\Delta}(x)=\frac{4}{3}\left[\frac{1+x^2}{1-x}
+\frac{2}{\log(x)}+(\frac{3}{2}-2\gamma_E)\delta(1-x)-(1-x)\right]\,,
\nonumber\\
& \displaystyle P_{\Delta g}(x)=6\left[\frac{x}{1-x}
+\frac{1}{\log(x)}+\frac{1-x}{x}+x(1-x)+(\frac{11}{12}-\frac{1}{18}n_f
-\gamma_E)\delta(1-x)\right]
\end{eqnarray}
and
\begin{eqnarray}
& \displaystyle \wt{D}_{\sto\ra\wth}(a,z,Q^2)\equiv\int_z^1{dy\over
y}D_{\sto\ra\wt{H}}(z/y,Q_0^2)
\frac{(-\log(y))^{(a\kappa-1)}}{\Gamma({a}\kappa)}\,,\nonumber\\
& \displaystyle \wt{D}_{g\ra\wth}(b,z,Q^2)\equiv \int_z^1{dy\over
y}D_{g\ra\wt{H}}(z/y,Q_0^2)
\frac{(-\log(y))^{(b\kappa-1)}}{\Gamma({b}\kappa)}\,,\nonumber\\
& \displaystyle\wt{\wt{D}}_{\sto\ra\wth}(z,Q^2)\equiv
\frac{1}{6-\frac{8}{3}}\int_{\frac{8}{3}}
^6{da}D_{\sto\ra\wt{H}}(a,z,Q^2)\,,\nonumber\\
&\displaystyle\wt{\wt{D}}_{g\ra\wth}(z,Q^2)\equiv
\frac{1}{6-\frac{8}{3}}\int_{\frac{8}{3}} ^6
dbD_{g\ra\wt{H}}(b,z/y,Q^2)\,,\nonumber
\end{eqnarray}
where $\kappa={6\over
{33-2n_f}}\log(\alpha_s(Q^2_0)/\alpha_s(Q^2))$, $\gamma_E$ is Euler
constant. Furthermore, at LL level we have the boundary condition
$D_{g\ra\wt{H}}(z/y,Q_0^2)=0$, thus the solution Eq.(\ref{ff-appr})
becomes
\begin{eqnarray}\label{ff-appr0}
&\displaystyle
D_{\sto\ra\wt{H}}(z,Q^2)=\wt{D}_{\sto\ra\wt{H}}(\frac{8}{3},z,Q^2) +
\kappa\int_z^1{dy\over
y}\wt{D}_{\sto\ra\wt{H}}(\frac{8}{3},z/y,Q^2)P_{\Delta}(y)
+O(\kappa^2)\,,\nonumber\\
&\displaystyle D_{g\ra\wt{H}}(z,Q^2)=\kappa\int_z^1{dy\over
y}\wt{\wt{D}}_{\sto\ra\wt{H}}(z/y,Q^2)P_{g\ra \sto}(y)
+O(\kappa^2)\,.
\end{eqnarray}
Numerically, it can be found that the first term for
$D_{\sto\ra\wt{H}}(z,Q^2)$ is much greater than the other terms in
the right hand side of the first equation of Eq.(\ref{ff-appr0}),
and then it is quite accurate to consider the first term only.
Moreover, due to the fact that the splitting function $P_{g \ra \sto
\bar{\sto}}$ must be suppressed greatly
$\cal{O}$(${m^2_Q/m_{\sto}^2}$) as $\sto$ is heavy $(m_{\sto}\geq
120\;{\rm GeV})$, we may safely conclude that
$D_{g\ra\wt{H}}(z,Q^2)\sim 0$ when $Q^2$ is not very great.
\begin{figure}
\centering
\includegraphics[scale=0.8]{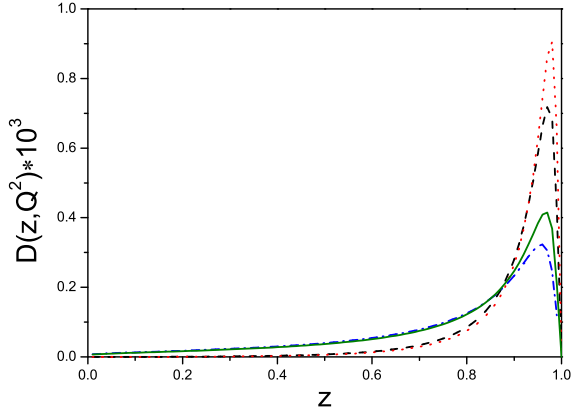}
\caption{Behavior of the fragmentation function (amplified by a
scale factor $10^3$ for convenience) for the light top-squark $\sto$
to $\wt{H}$ (here $\wt{H}$ precisely is the $S$-wave ($\sto\bar{b}$)
superhadron). The dotted and dashed lines are those stand for
Eq.(\ref{Dst}), the `initial' fragmentation function, with
$m_{\sto}=150$ GeV and $m_{\sto}=120$ GeV respectively. The solid
and dash-dot lines are those stand for the fragmentation function
evolving to the energy scale $Q=2$ TeV (a typical energy scale) with
$m_{\sto}=150$ GeV and $m_{\sto}=120$ GeV respectively.}
\label{figfrev}
\end{figure}

To precisely see the general behavior of the fragmentation function
obtained, we draw its curves in FIG.\ref{figfrev}. It can be found
that when the mass of the top-squark becomes heavier, the peak of
the curve for the fragmentation function increase higher
accordingly.

Furthermore, to see the characters of the obtained fragmentation
function, let us compare it with those for quarks ($Q$). The
fragmentation function for an anti-quark $\bar{Q}$ into a double
heavy meson ($\bar{Q}Q'$), e.g., a bottom anti-quark $\bar{b}$ into
$B_c$, which can be found in
Refs.\cite{chang92,chang921,chang922,chang923,braatenc,braatenc1,prod3}:
\begin{eqnarray}\label{fragcomp}
D_{\bar{b}}(z)=F_{\bar{b}}\cdot f_{\bar{b}}(z),
\end{eqnarray}
where $F_{\bar{b}}=\frac{8\alpha^2_s|\psi_0(0)|^2}{27Mm_c^2}$ and
\begin{eqnarray}\label{frag00}
f_{\bar{b}}(z)&=&\frac{z(1-z)^2}{(1-\lambda_1 z)^6}
\Bigg\{\left[12\lambda_2z-3(\lambda_1-\lambda_2)(1-\lambda_1z)
(2-z)\right](1-\lambda_1 z)z\nonumber\\
&& +6(1+\lambda_2z)^2(1-\lambda_1z)^2-
8\lambda_1\lambda_2z^2(1-z)\Bigg\} \,,
\end{eqnarray}
where $\psi_0(0)$ is wavefunction at origin of $B_c$,
$\lambda_1={m_b\over M}$, $\lambda_2={m_c\over M}$ and $M=m_b+m_c$.

%%frag function comparision
\begin{figure}
\centering
\includegraphics[scale=0.8]{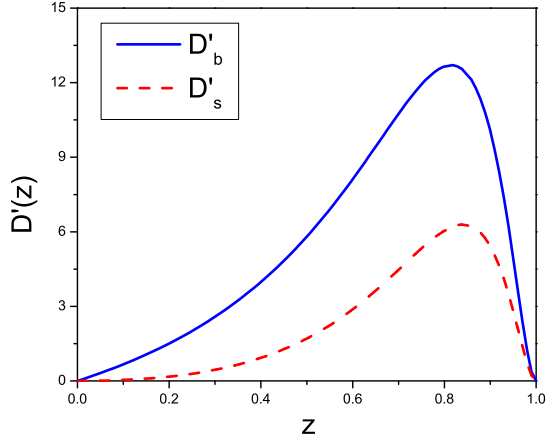}%
\caption{The comparison of the two types of fragmentation functions.
The upper curve (the solid one) is $D'_b(z)$, and the lower curve
(the dashed one) is $D'_s(z)$. The relevant parameters are taken as
$m_{\sto}=m_b=5.18$ GeV, $m_c=1.84$ GeV. The definitions and the
artificial assumption are taken as those in text.}
\label{figfragcomp} %\vspace{-0mm}
\end{figure}

To highlight the difference between the two types of fragmentation
functions, we remove the irrelevant factors $F_{\bar{b}}$ and
$F_{\sto}$ away and introduce the functions $D'_b(z)$ and $D'_s(z)$:
\begin{eqnarray}
&D'_b(z) = f_{\bar{b}}(z)\,\;\;{\rm and}\;\;\; D'_s(z) =
f_{\sto}(z)\,. \nonumber
\end{eqnarray}
One may see the differences clearly in asymptotic behaviors of the
two kinds of fragmentation functions that for $D'_b(z)$ and
$D'_s(z)$ they are $z$ and $z^2$ as $z\ra 0$ respectively; and the
same $(1-z)^2$ as $z\ra 1$. FIG.\ref{figfragcomp} depicts the two
kinds of fragmentation functions quantitatively. In the figure, the
function $D'_b(z)$ is taken precisely as the fragmentation function
for the $\bar{b}$ quark fragmenting into $S$-wave pseudoscalar state
of a double heavy meson $B_c$ or $B^*_c$, while the function
$D'_{\sto}(z)$ is the fragmentation function for the light
top-squark $\sto$ fragmenting into $S$-wave superhadron
$\wth=(\sto\bar{c})$. To contrast with the difference between these
two kinds of fragmentation functions, in FIG.\ref{figfragcomp} we
have assumed $m_{\sto}=m_{b}\simeq 5.18$ GeV artificially.

\section{Production of the superhadron at hadronic colliders}

Here we are adopting the fragmentation approach to estimate the
production of the superhadrons at hadronic colliders. According to
the NRQCD factorization theorem, the cross section of
$\wth$-production by collision of hadrons $H_1$ and $H_2$,
$d\sigma_{H_1H_2\to \wth X}$, can be factorized into three factors
as below:
\begin{eqnarray}\label{factD}
d\sigma_{H_1H_2\to \wth X} &=& \sum_{ijk} \int dx_1 \int dx_2 \int
dz f_{i/H_1}(x_1,
\mu_f) f_{j/H_2}(x_2, \mu_f) \nonumber\\
 && \cdot d\hat\sigma_{ij\to
k X}(x_1,x_2,z; \mu_f,\mu_R) \cdot D_{k\to \wt{H}}(z, \mu_{f}),
\end{eqnarray}
where $i,j$ and $k$ are parton species; $\mu_f$ corresponds to the
energy scale where the factorization is made; $\mu_R$ is the
renormalization energy scale for the hard subprocess;
$d\hat\sigma_{ij\to k X}(x_1,x_2,z; \mu_f,\mu_R)$ is the cross
section for the `hard subprocess' $ij\to k X$;
$D_{k\ra\wth}(z,\mu_{f})$ is the fragmentation function of `parton'
$k$ to $\wth$; $f_{i/H_1}(x_1, \mu_f)$ and $f_{j/H_2}(x_2, \mu_f)$
are the parton distribution functions (PDFs) in collision hadrons
$H_1$ and $H_2$ respectively. In this paper, as in most pQCD
calculations, we choose $\mu_f=\mu_R \sim
\sqrt{m^2_{\wt{H}}+p_T^2}\,$ \footnote{One will see later on that
the cross-section of the production decrease with $p_T$ rapidly.
Thus at Tevatron and LHC $\mu_f=\mu_R \sim
\sqrt{m^2_{\wt{H}}+p_T^2}$ is not high enough that $m_{\sto}$ can be
considered as zero.}, and, as a consequence of the choice, we may
further set $D_{g\ra\wt{H}}(z,Q^2)=0$ quite safely as argued above.
Therefore, $k$ in Eq.(\ref{factD}) `runs over' $\sto$ only, i.e.
$k=\sto$.

By naive considerations, of all the possible hard subprocesses for
the production ($ij\to k X\,,\; k=\sto$), the gluon-gluon fusion
$g+g\rightarrow \wt{t}_1 +\bar{ \wt{t}}_1$ and the quark-antiquark
annihilation $q + \bar{q} \rightarrow \wt{t}_1 +\bar{\wt{t}}_1$
(here $q$ and $\bar{q}$ are light quarks) are in the same order of
strong coupling $\alpha_s$, so they may be the most important ones
for the production at the colliders: Tevatron and LHC. The gluon
component of PDFs in small $x$ region is the greatest, so the
gluon-gluon fusion should be the most important one at LHC; whereas,
at Tevatron, due to a comparatively low CM energy, the
energy-momentum fraction $x$ of the gluon parton must be big enough
to produce the $\sto\bar{\sto}$ pair, so as the case of top-quark
pair production at Tevatron probably the components of valance
quarks, instead of the gluon, play more important role (a review of
this point can be found in Ref.\cite{top}). Therefore, first of all
we highlight these two subprocesses for the production. Note that
according to Ref.\cite{s-beenmix} the production of the top-squark
pair $\sto\bar{\wt{t}_2}$ or $\wt{t}_2\bar{\sto}$ at the hadronic
colliders is small, so we do not take them into account.

%gg->st+st*
\begin{figure}
\centering
\includegraphics[scale=0.9]{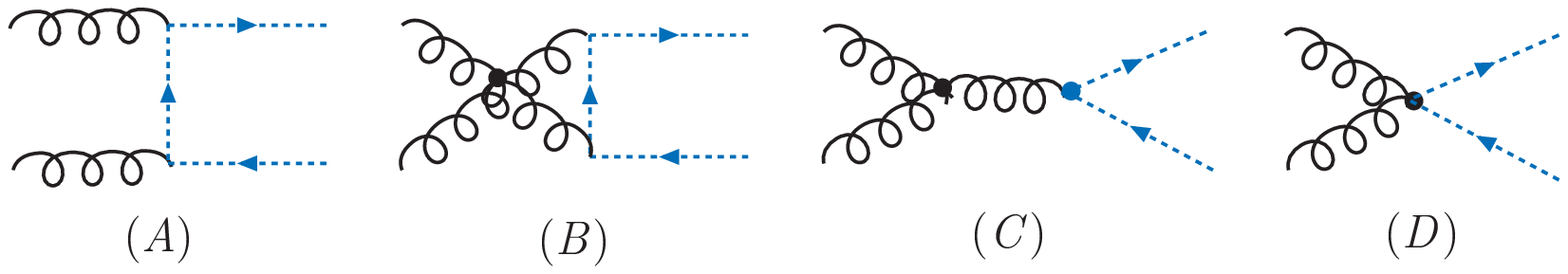}
\caption{The lowest order Feynman diagrams for the gluon-gluon
fusion subprocess $g + g \ra \sto + \bar{\wt{t}}_1$.}
\label{figggstop}
\end{figure}

Now let us calculate the gluon-gluon fusion subprocesses first. To
the lowest order (tree level), there are four Feynman diagrams as
shown in FIG.\ref{figggstop}. The corresponding amplitudes read
\begin{eqnarray}
M_A&=&g_s^2T^{ab}\frac{4p_{1}^{\mu}p_{2}^{\nu}\e_{\mu}(k_1)\e_{\nu}(k_2)}
      {(p_1-k_1)^2-m_{\sto}^2},\\
M_B&=&g_s^2T^{ba}\frac{4p_{2}^{\mu}p_{1}^{\nu}\e_{\mu}(k_1)\e_{\nu}(k_2)}
      {(p_1-k_2)^2-m_{\sto}^2},\\
M_C&=&g_s^2(T^{ab}-T^{ba})\frac{
(k_{2}-k_1)^{\mu'}g^{\mu\nu}-2k_{2}^{\mu}g^{\mu'\nu} +
2k_{1}^{\nu}g^{\mu'\mu}}{(k_1+k_2)^2}(p_1-p_2)_{\mu'}\e_{\mu}(k_1)\e_{\nu}(k_2),\\
M_D&=&g_s^2(T^{ab}+T^{ba})\e_{\mu}(k_1)\e_{\nu}(k_2)g^{\mu\nu}.
\end{eqnarray}
where $\e$ is the polarization vector of gluon. When taking the
axial gauge with a fixed four-vector $n$, the summation of
polarization vector reads
\begin{displaymath}
\sum_{\lambda}\e^*_{\mu}(k,\lambda)\e_{\nu}(k,\lambda)=-g_{\mu\nu}
-\frac{k_{\mu}k_{\nu}n^2}{(k\cdot
n)^2}+\frac{k_{\mu}n_{\nu}+k_{\nu}n_{\mu}}{k\cdot n}.
\end{displaymath}

The differential cross section is
\begin{eqnarray}\label{ggcs0}
d\hat\sigma(gg\to \sto\wt{\bar{t}}_1 )&=&
\frac{3\pi^2\alpha_s^2}{16\pi\hat{s}^2}
\left[1-2A-\frac{1}{9}\right] \left[1-2\frac{m^2_{\sto}}{A\hat{s}}
(1-\frac{m^2_{\sto}}{A\hat{s}})\right]d\hat{t},
\end{eqnarray}
where $A=(\hat{t}-m_{\sto}^2)(\hat{u}-m^2_{\sto})/\hat{s}^2$.
$\hat{s}$, $\hat{t}$ and $\hat{u}$ are Mandelstam variables of the
subprocess,
\begin{eqnarray}
\label{mans}\hat{s}&=&(k_1+k_2)^2,\\
\label{mant}\hat{t}&=&(p_1-k_1)^2,\\
\label{manu}\hat{u}&=&(p_1-k_2)^2,
\end{eqnarray}
which satisify $\hat{s}+\hat{u}+\hat{t}=2m^2_{\sto}$.

For the quark-antiquark annihilation subprocess, to the lowest order
there is only one Feynman diagram and its Feynman amplitude reads:
\begin{eqnarray}
M_{q\bar{q}\rightarrow
\wt{t}_1\bar{\wt{t}_1}}&=&g_s^2T^{aa}\frac{(p_1^{\mu}
-p_2^{\mu})\bar{v}(k_2)\gamma_{\mu}u(k_1)} {(k_1+k_2)^2}\ ,
\end{eqnarray}
where $k_1$ and $k_2$ are the four momenta for the quark and
antiquark respectively. The differential cross section is obtained
as below
\begin{eqnarray}\label{qqcs1}
d\hat\sigma(q\bar{q}\to \sto\wt{\bar{t}}_1)&=&
\frac{\pi\alpha_s^2[\hat{s}^2-4\hat{s}
m^2_{\sto}-(\hat{t}-\hat{u})^2]}{9\hat{s}^{4}}d\hat{t} ,
\end{eqnarray}
where the Mandelstam variables are defined as
Eqs.(\ref{mans},\ref{mant},\ref{manu}). To calculate the production
via the subprocess of quark-antiquark annihilation, as stated above,
we are interested in seeing the valance quarks' contributions,
especially, the production at Tevatron, so here we are considering
the contributions only from the light quarks in PDFs to the
production. In nucleons only the light quarks $u$ and $d$ may be
their valance quarks.

The total hadronic cross section is calculated via the two
subprocesses in terms of the factorization formulation
Eq.(\ref{factD}) and with the help of the fragmentation function
Eq.(\ref{Dst}). The differential cross-section for the gluon-gluon
fusion is in terms of Eq.(\ref{ggcs0}) and the differential
cross-section for the quark-antiquark annihilation is in terms of
Eq.(\ref{qqcs1}). Since the present calculations are at the lowest
order only, so the version CTEQ6L \cite{6lcteq} for the parton
distribution functions (PDFs) is taken, and for definiteness, we
take $m_b=5.18$ GeV, $m_c=1.84$ GeV and assume two possible values
for $m_{\sto}$: $m_{\sto}=120$ or $150$ GeV. In addition, the
parameter $\Lambda_{QCD}$ in the running coupling constant
$\alpha_s$ is taken as $0.216$ GeV. The energy scale for the QCD
factorization formulas is chosen as the `transverse mass' of the
produced superhadron: $\sqrt{m^2_{\wt{H}}+p_T^2}$.

\begin{table}
\begin{center}
\caption{Hadronic cross sections (in unit: fb) for the superhadrons
($\sto\bar{c}$) and ($\sto\bar{b}$) with $J^P=({1\over 2})^-$. The
parameters appearing in the estimate are taken as those stated in
text.}\vspace{2mm}
\begin{tabular}{|c|c|c|c|c|c|}
\hline \multicolumn{2}{|c|}{} &
\multicolumn{2}{|c|}{~LHC~($\sqrt{S}$=14. TeV)~}
&\multicolumn{2}{|c|}{TEVATRON~($\sqrt{S}=1.96$ TeV)}\\
\hline %\hline
\multicolumn{2}{|c|}{~Constituents~ }& ~subprocess $gg$~ &
subprocess $q\bar{q}$
& ~~~subprocess $gg$~~~ & subprocess $q\bar{q}$  \\
\hline\hline
$m_{\sto}=120$&($\sto\bar{c}$) & 114.51 & 0.36469 & 0.26975 &1.2E-3\\
\cline{2-6}
           GeV         &($\sto\bar{b}$) & 30.489 & 0.10374 & 0.0696 & 3.E-4 \\
\hline\hline
$m_{\sto}=150$ &($\sto\bar{c}$)  & 42.176 & 0.14591 & 0.0537 & 2.E-4 \\
\cline{2-6}
          GeV         &($\sto\bar{b}$) & 11.812 & 0.0431 & 0.0142 & 7.E-5 \\
\hline
\end{tabular}
\label{tabcs} \vspace{-6mm}
\end{center}
\end{table}

\begin{figure}
\centering
\includegraphics[width=0.45\textwidth]{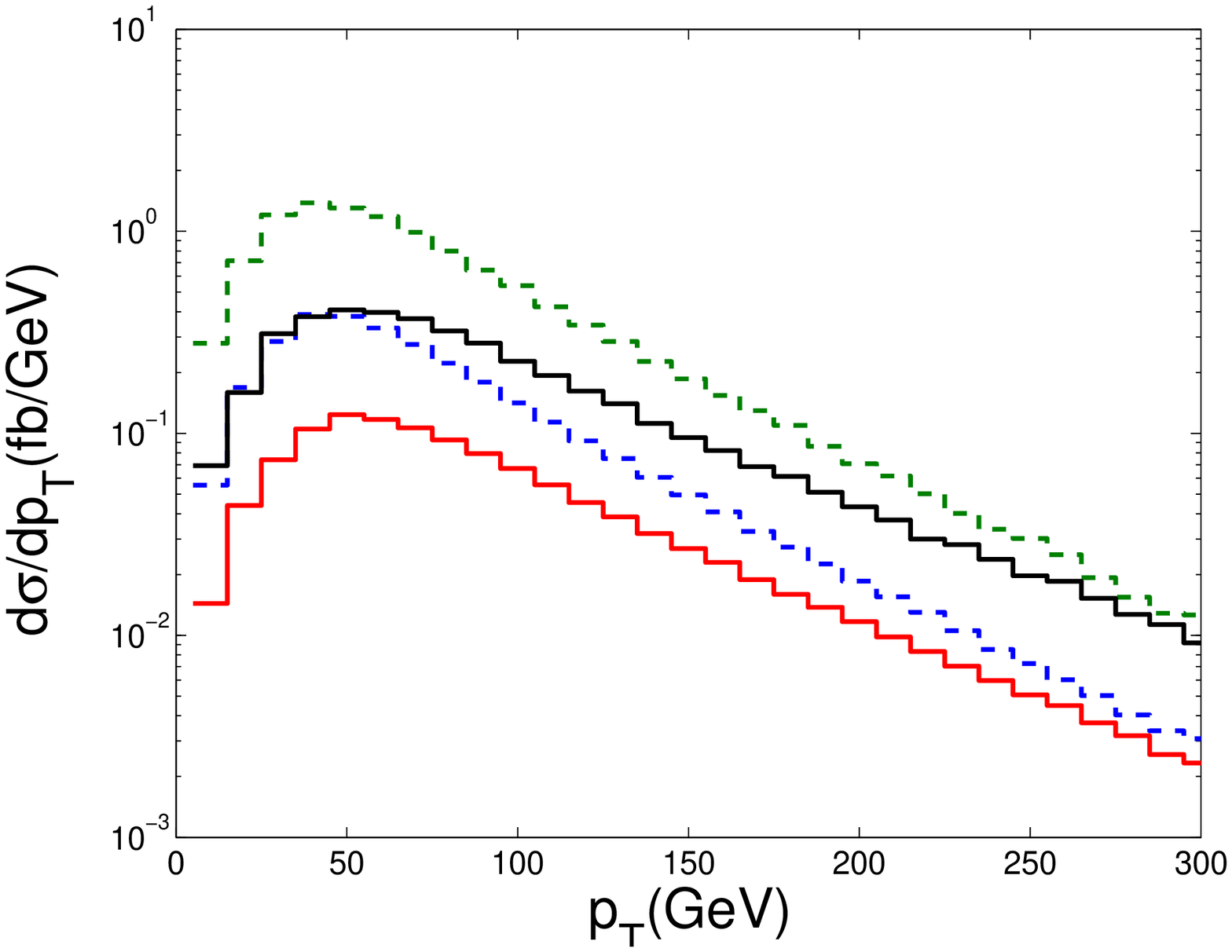}%
\hspace{0.2cm}
\includegraphics[width=0.45\textwidth]{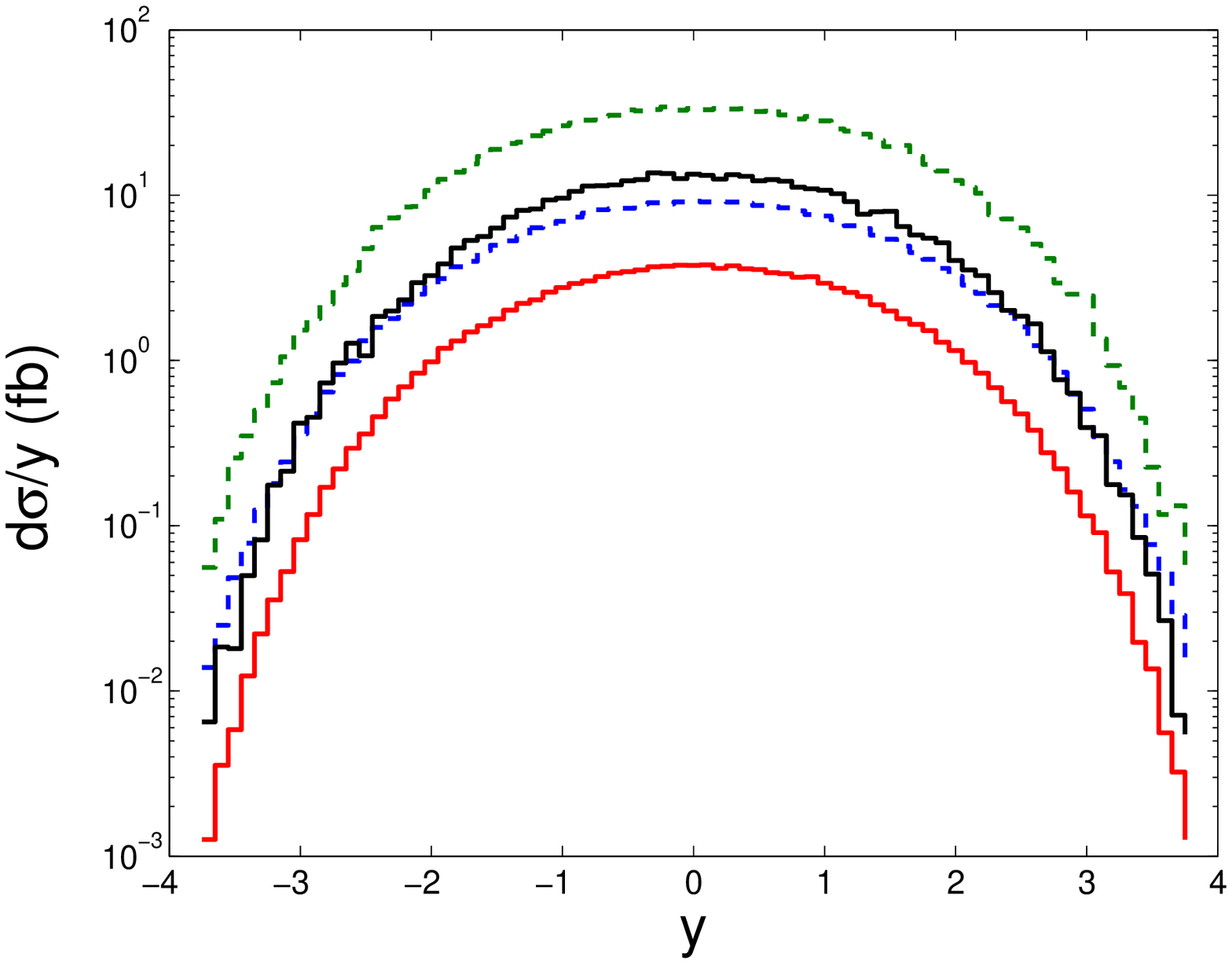}
\caption{The distributions of the transverse momentum $P_T$ (left
figure) and rapidity $y$ (right figure) for the superhadron $\wt{H}$
produced via gluon-gluon fusion at LHC. For $P_T$ distribution, a
rapidity cut $|y|<1.5$ is made. The upper one of the two dash lines
corresponds to the distribution for the superhadron ($\sto\bar{c}$)
production with $m_{\sto}=120$ GeV being assumed, the lower one to
that for the superhadron ($\sto\bar{b}$) production with
$m_{\sto}=120$ GeV; The upper one of the two solid lines to the
distribution for the superhadron ($\sto\bar{c}$) production with
$m_{\sto}=150$ GeV, the lower one to the distribution for the
superhadron ($\sto\bar{b}$) production with $m_{\sto}=150$ GeV.}
\label{lhcgg} \vspace{-0mm}
\end{figure}

\begin{figure}
\centering
\includegraphics[width=0.45\textwidth]{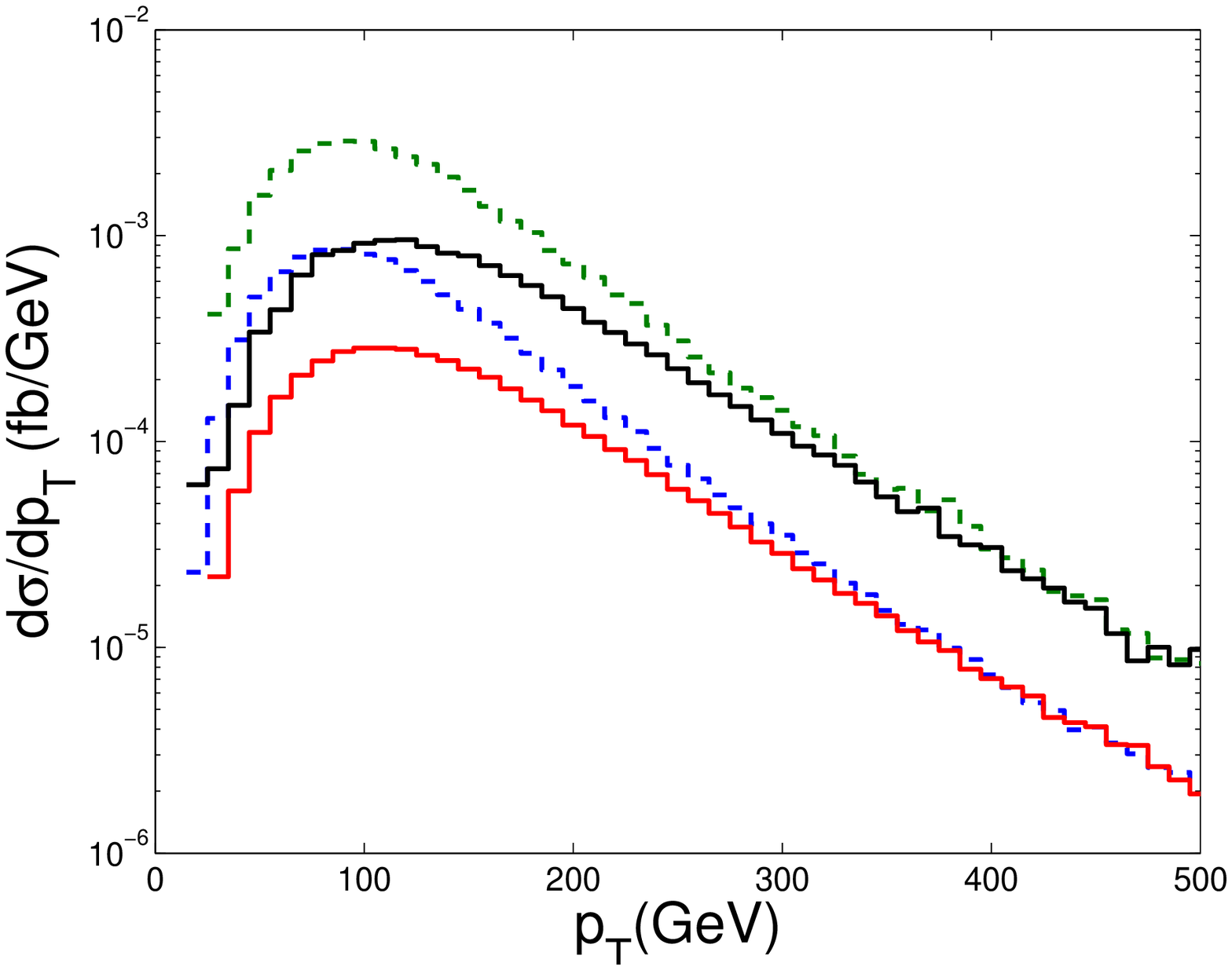}%
\hspace{0.2cm}
\includegraphics[width=0.45\textwidth]{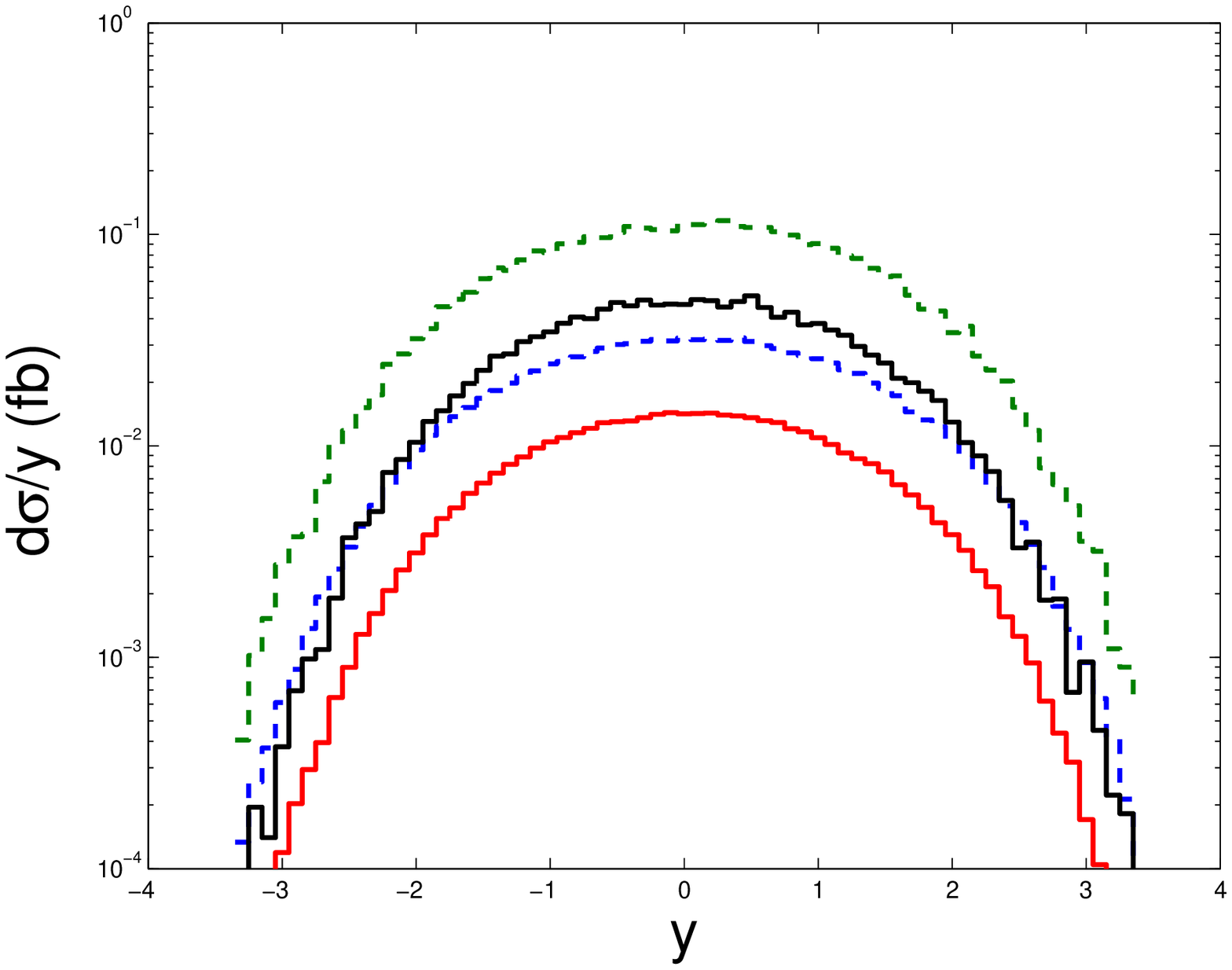}
\caption{The distributions of the transverse momentum $P_T$ (left
figure) and rapidity $y$ (right figure) for the superhadron $\wt{H}$
produced via quark-antiquark annihilation at LHC. For $P_T$
distribution, a rapidity cut $|y|<1.5$ is made. The upper one of the
two dash lines corresponds to the distribution for the superhadron
($\sto\bar{c}$) production with $m_{\sto}=120$ GeV being assumed,
the lower one to that for the superhadron ($\sto\bar{b}$) production
with $m_{\sto}=120$ GeV; The upper one of the two solid lines to the
distribution for the superhadron ($\sto\bar{c}$) production with
$m_{\sto}=150$ GeV, the lower one to the distribution for the
superhadron ($\sto\bar{b}$) production with $m_{\sto}=150$ GeV.}
\label{lhcqq} \vspace{-0mm}
\end{figure}

\begin{figure}
\centering
\includegraphics[width=0.45\textwidth]{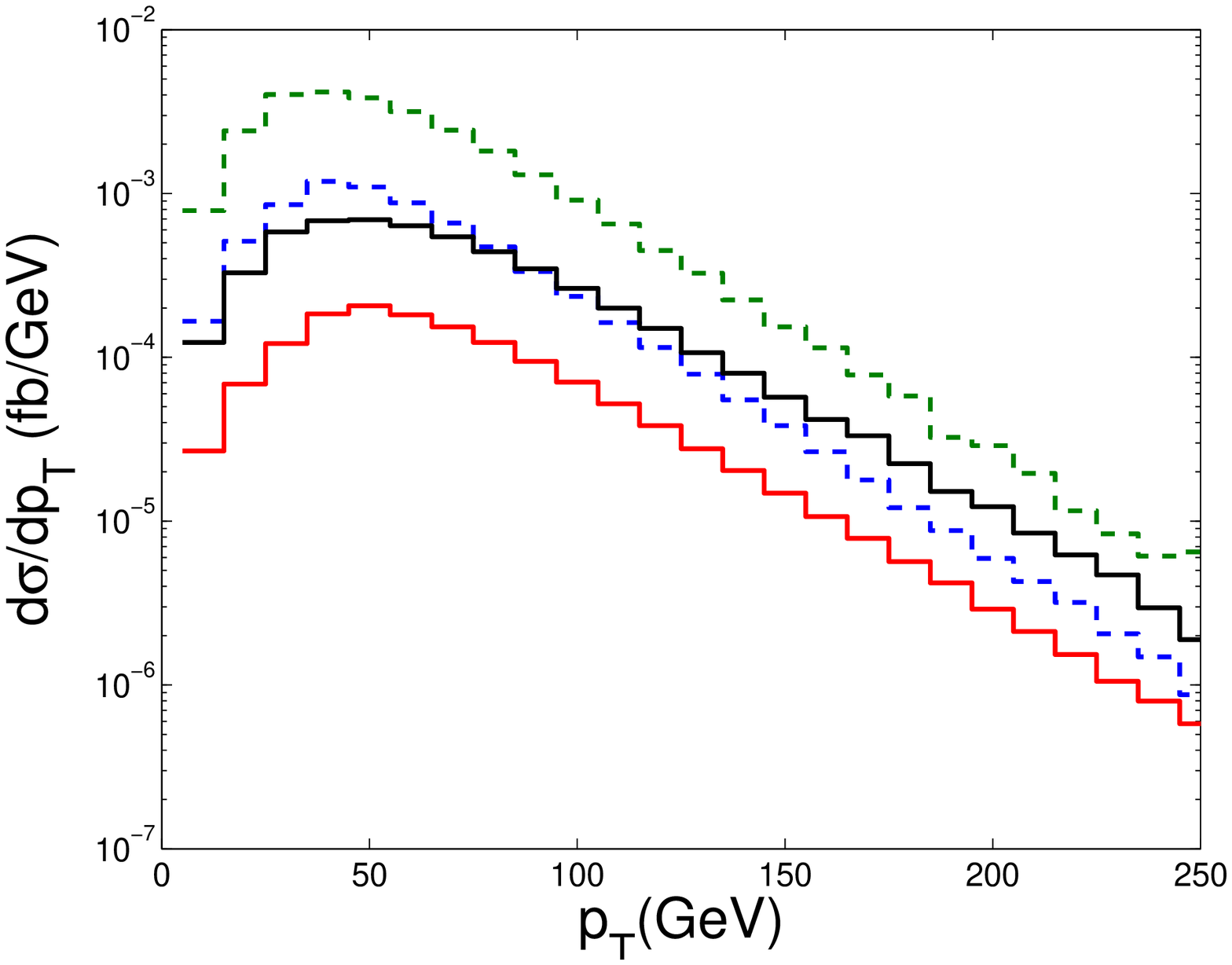}%
\hspace{0.2cm}
\includegraphics[width=0.45\textwidth]{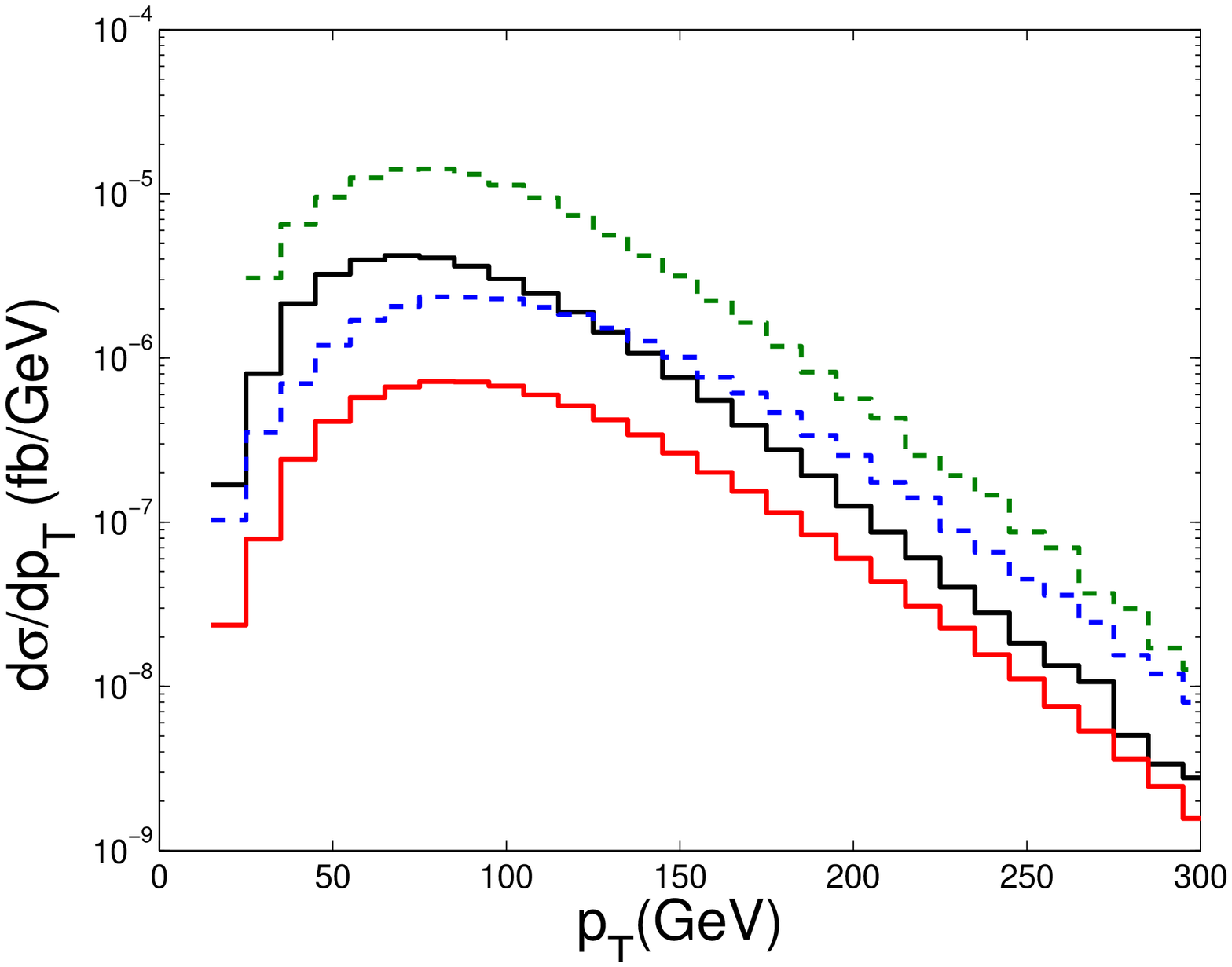}
\caption{The $P_T$ distributions of the superhadrons $\wt{H}$
produced at Tevatron via the gluon-gluon fusion (left figure) and
the quark-antiquark annihilation (right figure) respectively. For
the distributions, a rapidity cut $|y|<0.6$ is made. The upper one
of the two dash lines corresponds to the distribution for the
superhadron ($\sto\bar{c}$) production with $m_{\sto}=120$ GeV, the
lower one to the distribution for the superhadron ($\sto\bar{b}$)
production with $m_{\sto}=120$ GeV; The upper one of the solid lines
to the distribution for the superhadron ($\sto\bar{c}$) production
with $m_{\sto}=150$ GeV, the lower one to the distribution for the
superhadron ($\sto\bar{b}$) production with $m_{\sto}=150$ GeV.}
\label{teva} \vspace{-0mm}
\end{figure}

The total hadronic cross sections obtained at Tevatron and LHC are
in TABLE \ref{tabcs}.  From the table one may see that the cross
section for hadronic production of the superhadron $\wth$ at
Tevatron is much smaller than that at LHC (almost by three order of
magnitude), so when $\sto$ possesses the behaviors as assumed here
and considering the final possible integrated luminosity, it is
hopeless to observe $\wth$ at Tevatron, but it is hopeful at LHC.
TABLE \ref{tabcs} also shows that the hadronic cross sections of the
superhadron $\wth$ at Tevatron and at LHC decrease as the mass of
the light scalar top quark $m_{\sto}$ increasing. Moreover the cross
sections for the superhadron production via gluon-gluon fusion are
much larger than those via the annihilation at Tevatron and at LHC
both. Hence, the contribution from quark-antiquark annihilation can
be ignored in comparison to the dominant contribution from the
gluon-gluon fusion.

To present more features of the production, we also draw curves to
show the distributions of the produced superhadron. The differential
cross-sections vs the transverse momentum $p_T$ and the rapidity $y$
of the produced superhadron via gluon-gluon fusion at LHC are drawn
in FIG.\ref{lhcgg}, while those via the quark-antiquark annihilation
are drawn in FIG.\ref{lhcqq}. The distributions of transverse
momentum $p_T$ for the superhadron production at Tevatron via
gluon-gluon fusion and quark-antiquark annihilation are drawn in
FIG.\ref{teva} respectively.

\section{Discussion and Conclusions}

Here the fragmentation function of the light top-squark $\sto$ to
heavy superhadrons ($\sto \bar{c}$) and ($\sto \bar{b}$) is reliably
computed as that in the case of a heavy quark to a double heavy
meson. To see the characteristics of the fragmentation function,
comparisons of the obtained fragmentation function for the light
top-squark with those for heavy quarks are made by drawing curves
with suitable parameters in FIG.\ref{figfragcomp}. When $z$
approaches to zero, the fragmentation functions for the top-squark
(a scalar particle) approach to zero as $z^2$, instead of those for
a heavy quark (a fermion particle) which behave as $z$; and both of
them have a similar asymptotic behavior when $z$ approaches to 1.

Using the obtained fragmentation function of the superhadron
$\wt{H}$ (either $\sto \bar{c}$ or $\sto \bar{b}$) and under the
fragmentation approach up-to leading logarithm, the cross sections
and $P_T$ ($y$) distributions for $\wt{H}$ have been computed at the
energies of Tevatron and LHC. In the computation, the gluon-gluon
fusion and light quark-antiquark annihilation as the hard subprocess
for the hadronic production of the superhadron $\wth$ are taken into
account precisely. When calculating $P_T$ distributions, different
rapidity cuts are taken, i.e. $|y|<1.5$ at LHC and $|y|<0.6$ at
Tevatron. From the cross sections and $P_T$ ($y$) distributions, one
may conclude that one cannot collect enough events for observing the
superhadron at the hadronic collider Tevatron, even if the
parameters of the supersymmetric model are in a very favored region.
In contrary, enough events for experimental observation of the
superhadrons can be produced (collected) without difficulty at the
forthcoming collider LHC. Namely if the coming `new physics' is
supersymmetric and the parameters are in the favored region of the
concerned superhadrons and allowed by all kinds of the existent
experimental observations, Tevatron is not a good `laboratory' to
observe the possible superhadron(s), while LHC may be a good one.
Moreover, it can be found from TABLE \ref{tabcs} that the production
cross-section via $q + \bar{q}\rightarrow \wt{t}_1 + \bar{\wt{t}}_1$
is much smaller than that via gluon-gluon fusion subprocess at
Tevatron and LHC. To compare with the top-quark production, let us
note here that the quark-antiquark annihilation for the top-quark
production is comparable to the gluon-gluon fusion at LHC
\cite{lhctop,lhctop1,lhctop2,lhctop3,lhctop4}, while for the so
heavy top-quark the quark-antiquark annihilation mechanism is
dominant over the gluon-gluon fusion at Tevatron
\cite{tevtop,tevtop1,tevtop2,tevtop3}. So the cross-sections for
single top production via $q+\bar{q}'\to t+\bar{b}$ \cite{qqbart}
and $q+b\to q'+t$ \cite{qbt}, which are comparable to the
quark-antiquark annihilation, are also important for the top-quark
production.

On the production of the superhadrons at Tevatron and at LHC, for
the reason precisely pointed out in the above section we have
highlighted the two mechanisms via the hard subprocesses:
gluon-gluon fusion and light quark-antiquark annihilation so far. In
fact, there may be some other mechanisms for producing the
superhadrons which may contribute greater than that via light
quark-antiquark annihilation and even so sizable to be comparable
with that via gluon-gluon fusion. For instance, when a comparatively
light chargino ($m_{\wt{\chi}^{\pm}}\leq {\cal{O}}$(TeV)) is allowed
in the same SUSY models, the `single top-squark production' such as
that via $g+b\to \sto+\wt{\chi}^{-}_{1/2}$ can be the case: its
contribution may be greater than that via light quark-antiquark
annihilation and even so sizable to be comparable with that via
gluon-gluon fusion. It is very similar to the top
production\cite{lhctop,lhctop1,lhctop2,lhctop3,lhctop4}: at the LHC
the single top production via the process $g+b \to W+t$ has a cross
section of about 60 pb while the cross section via gluon-gluon
fusion $g+g\to t + \bar{t}$ is roughly about 760 pb and the cross
section via quark-antiquark annihilation $q+\bar{q} \to t + \bar{t}$
is roughly about 40 pb. Moreover if the bottom-squark $\wt{b_1}$ is
also comparatively light ($m_{\wt{b_1}}\leq {\cal{O}}$(TeV)) in the
same SUSY models, then the production via $q+\bar{q'}\to
\sto+\bar{\wt{b_1}}$ can be quite great too. However, all of the
possibilities depends on the parameters of the relevant SUSY models,
thus we would not calculate them precisely in the paper. As for the
production via the subprocesses such as the annihilation of
top-quark and anti-top-quark through gluino $\wt{g}$ or photino
$\wt{\gamma}$ exchanging $t+\bar{t}\ra \sto+\bar{\sto}$, and a
top-quark `scattering' on a gluon: $g+t\ra \sto+\wt{g}(\wt{\gamma})$
etc, we are sure that their contributions to the superhadron
production are very tiny due to the smallness of the PDF of
top-quark in the colliding hadrons. Anyway, for the accuracy of the
present estimate and such a `light' top-squark $m_{\sto}=120\sim
150$GeV, the contribution via the quark-antiquark annihilation hard
subprocess to the production can be negligible in comparison to the
dominant gluon-gluon fusion mechanism at Tevatron and LHC both, that
is quite different from the top-quark case.

Since the decay of the heavy superhadrons $\wth$ i.e.
$(\sto\bar{Q})$ with $Q=c, b$ is via the light top-squark $\sto$ or
via the involved heavy quark $\bar{Q}$ with proper relative decay
possibility, so there are two typical decay channels for $\wth$: one
is the decay of the light top-squark $\sto$ with the heavy quark
$\bar{Q}$ acting as a `spectator' and the other is the decay of the
heavy quark $\bar{Q}$ with the light top-squark $\sto$ acting as a
`spectator'. The second decay channel may be quite different from
that of the decay for a light top-squark $\sto$ itself, and then it
shall present certain characteristics. Therefore, we think that in
order to observe and identify (discover) the light top squark $\sto$
experimentally, one may try to gain some advantages via observing
the characteristics of the heavy superhadrons $\wth$ decay.

If the heavy superhadrons are really observed experimentally, it
will be a good news not only for the relevant SUSY model(s) but also
for the QCD-inspired potential model, because it will open a fresh
field, i.e. the potential model will need to be extended to treat
systems with binding of a fermion and a scalar boson.

As it is known, the fragmentation of a heavy quark $b$ or $c$ to a
double heavy meson $\eta_b$ or $\eta_c$ is quite smaller than that
of the heavy quark to a heavy meson $B$ or $D$ i.e. with a relative
possibility about $10^{-4}\sim 10^{-3}$
\cite{chang92,chang921,chang922,chang923,braatenc,braatenc1,prod3},
thus with the same reason one may be quite sure that the
fragmentation function of top-squark $\sto$ to light superhadrons
($\sto\bar{q}$), $q=u, d, s$ are much greater than that of
top-squark $\sto$ to heavy superhadrons ($\sto\bar{Q}$) $Q=c, b$.
Namely, by conjecture, the fragmentation function of top-squark
$\sto$ to light superhadrons ($\sto\bar{q}$) can be about
($10^{3}\sim 10^{4}$) of the top-squark $\sto$ to the heavy
superhadron ($\sto\bar{Q}$) one. With such an enhancement this
large, the light superhadrons may be produced numerously, and then
one may collect enough events for experimental observation even at
Tevatron. However without the additional characteristics due to the
decay heavy quark $b$ or $c$ of the heavy superhadrons, the light
superhadrons may be comparatively difficult to be identified
experimentally.

Finally, we should note here that the computation and discussion in
the paper are explicitly based on the assumption that the light
color-triplet top-squark does exist in certain SUSY models. As a
matter of fact, the results in the present paper are true for a
variety of the SUSY models, in which even the light top-squark
$\sto$ is not the lightest SUSY object in nontrivial color. As long
as in the SUSY models concerned, the lightest SUSY partner is a
scalar in color triplet and has a lifetime long enough to form
hadrons before decaying, our results as presented here remain
meaningful by simply replacing the light top-squark $\sto$ with the
corresponding lightest SUSY partner.

\hspace{3cm}

\noindent {\Large\bf Acknowledgement:} This work was supported
partly by the Natural Science Foundation of China (NSFC). The
authors would like to thank J.M. Yang and J.P. Ma for helpful
discussions.

%%%=====================================================

\end{document}